\documentclass[%
prl,%
preprint,
aps,%
showpacs,%
a4paper,%
superscriptaddress,%
floatfix%
]{revtex4-1}

\usepackage{graphicx}
\usepackage{bm}
\usepackage{amsmath}
\usepackage{amssymb}
\usepackage{float}
\usepackage[utf8]{inputenc}
\usepackage[T1]{fontenc}
\usepackage[USenglish]{babel}
\usepackage{microtype}
\usepackage{nicefrac}
\usepackage{xspace}
\usepackage{subcaption}
\usepackage[color=black!10]{todonotes}

\usepackage[colorlinks,citecolor=blue,urlcolor=blue,linkcolor=blue,bookmarks=false,filecolor=blue,runcolor=blue]{hyperref}
\usepackage[all]{hypcap}
\newcommand{\figref}[1]{Fig.~\ref{#1}}

\newcommand{\unit}[1]{~\mathrm{#1}}

\renewcommand{\Im}{%
\mathrm{Im}
}
\newcommand{\GX}{$\Gamma\text{X}$\xspace}
\newcommand{\GY}{$\Gamma\text{Y}$\xspace}
\newcommand{\rangstrom}[1]{
\ensuremath{#1\,\text{\AA}^{-1}}
}
\newcommand{\bp}{BP\xspace}
\begin{document}
\title{Anisotropic Particle-Hole Excitations in Black Phosphorus}
\author{R. Schuster}
\author{J. Trinckauf}
\author{M. Knupfer}
\affiliation{IFW Dresden, Institute for Solid State Research, P.O. Box 270116, D-01171 Dresden, Germany}
\author{B. Büchner}
\affiliation{IFW Dresden, Institute for Solid State Research, P.O. Box 270116, D-01171 Dresden, Germany}
\affiliation{Department of Physics, Technische Universität Dresden, 01062 Dresden, Germany}

\date{\today}

\begin{abstract} 
We report about the energy and momentum resolved optical response of black phosphorus (\bp) in its bulk form. Along the armchair direction of the puckered layers we find a highly dispersive mode that is strongly suppressed in the perpendicular (zig-zag) direction. This mode emerges out of the single-particle continuum for finite values of momentum and is therefore interpreted as an exciton. We argue that this exciton, which has already been predicted theoretically for phosphorene – the monolayer form of \bp – can be detected by conventional optical spectroscopy in the two-dimensional case and might pave the way for optoelectronic applications of this emerging material. 
\end{abstract}
\maketitle

%\section{Introduction}
Among the different materials which are currently under investigation in the “post-graphene” era, black phosphorus (\bp) in its monolayer form phosphorene is of particular interest \cite{Churchill_NatNano_2014_v9_p330}. Due to its intrinsic band-gap of about $0.3\unit{eV}$ in the bulk and much higher values for the monolayer  \cite{Tran_Phys.Rev.B_2014_v89_p235319,Das_NanoLett._2014_v14_p5733}, devices built from \bp show substantially increased performance compared to pristine graphene in terms of e.g. on/off ratios \cite{Li_NatNano_2014_v9_p372,Das_NanoLett._2014_v14_p5733}. 

In addition, very high mobility values (up to $10^4\unit{cm}^2\unit{V}^{-1}\unit{s}^{-1}$) have been predicted theoretically for phosphorene \cite{Qiao_NatCommun_2014_v5_p4475} and at least to some extend confirmed experimentally \cite{Li_NatNano_2014_v9_p372}. What is particularly remarkable is a strong in-plane anisotropy of the transport and optical properties \cite{Qiao_NatCommun_2014_v5_p4475,Xia_NatCommun_2014_v5_p4458,Liu_ACSNano_2014_v8_p4033,Cai_ScientificReports_2014_v4_p6677,Zhang_ACSNano_2014_v8_p9590} that is, however, well known already for the bulk \cite{Narita_PhysicaB+C_1983_v117_118Part1_p422,Maruyama_Bull.Chem.Soc.Jpn._1986_v59_p1067}. Even more, it has been predicted that phosphorene allows for the fabrication of fullerene- and nanotube-like aggregates \cite{Guan_Phys.Rev.Lett._2014_v113_p226801} with highly tunable electronic properties. Together with the known tendency to undergo several structural phase transitions under pressure – which eventually results in a superconducting ground state with $T_c\approx 10\unit{K}$ \cite{Kawamura_SolidStateCommun._1984_v49_p879} – this promotes \bp to a versatile platform for possible applications.

While a thorough understanding of the charge response is of utmost importance for all optical applications, there is a surprising lack of corresponding data, in particular in the near-infrared to visible regime.

Here, by employing electron energy-loss spectroscopy (EELS) in transmission, we bridge this gap. We find strongly anisotropic optical properties which are the result of an excitonic state that is strongly polarized  and – at least in the bulk of \bp – exists only away from the center of the Brillouin zone (BZ).

%\section{Experiments and Results}
EELS in transmission is a bulk-sensitive scattering technique whose cross-section is proportional to the so called loss-function $L(\bm{q},\omega)=\Im(-1/\epsilon(\bm{q},\omega))$, with $\epsilon(\bm{q},\omega)$ the momentum- and energy-resolved dielectric function \cite{Sturm_Z.Naturforsch._1993}. It has been successfully applied to investigate collective charge modes in numerous condensed-matter systems (see \cite{Roth_J.ElectronSpectrosc.Relat.Phenom._2014_v195_p85,Fink_J.ElectronSpectrosc.Relat.Phenom._2001_v117-118_p287} for an overview). In particular, it has proven useful for the investigation of excitonic states in organic semiconductors \cite{Schuster_Phys.Rev.Lett._2007_v98_p37402,Roth_J.Chem.Phys._2012_v136_p204708,Roth_NewJ.Phys._2013_v15_p125024}.  In addition it is possible to retrieve the complete dielectric function $\epsilon(\bm{q},\omega)=\epsilon_1(\bm{q},\omega)+\mathrm{i}\epsilon_2(\bm{q},\omega)$ via a Kramers-Kronig analysis (KKA) of the measured data.

For the present experiments thin films have been prepared via exfoliation from a single crystal purchased from “2d Semiconductors Inc.”\footnote{http://www.2dsemiconductors.com/}. The films ($d\approx100-200\unit{nm}$) were put on standard electon-microscopy grids and then transferred to the spectrometer where they have been aligned \emph{in-situ}  with electron diffraction showing the high quality of our samples and allowing for polarization dependent investigations along well defined directions within the puckered layers of \bp. The measurements were carried out using a purpose-built transmission electron energy-loss spectrometer \cite{Fink_Adv.Electron.ElectronPhys._1989} with a primary electron energy of $172\unit{keV}$ and energy and momentum resolutions of $\Delta E = 80\unit{meV}$ and $\Delta q =\rangstrom{0.035}$, respectively and at temperatures of $T\approx20\unit{K}$ to minimize thermal broadening. All discussed features remain qualitatively equivalent at room temperature.

First-principles calculations have been performed with the FPLO package \cite{Koepernik_Phys.Rev.B_1999_v59_p1743}.
We employed the crystal structure given in \cite{Brown_ActaCryst._1965_v19_p684} except for the in-plane lattice constants which we determined experimentally from electron diffraction to be 4.363\AA\ along the armchair and 3.206\AA\ along the zig-zag direction, respectively. The total density was converged on a grid of $12\times12\times12$ irreducible $k$-points. The generalized gradient approximation (GGA) for the exchange correlation potential has been used as parametrized by Perdew, Burke and Ernzerhof \cite{Perdew_Phys.Rev.Lett._1996_v77_p3865}.

Under ambient conditions \bp in its bulk form adopts an orthorhombic lattice with puckered layers spanned by the $x$- and $y$-axes along which the atoms are arranged in armchair and zig-zag coordination, respectively. This symmetry allows for two independent in-plane components of the dielectric tensor $\epsilon_{xy}(\bm{q},\omega)$. 

\begin{figure*}[ht]
\includegraphics[scale=.5]{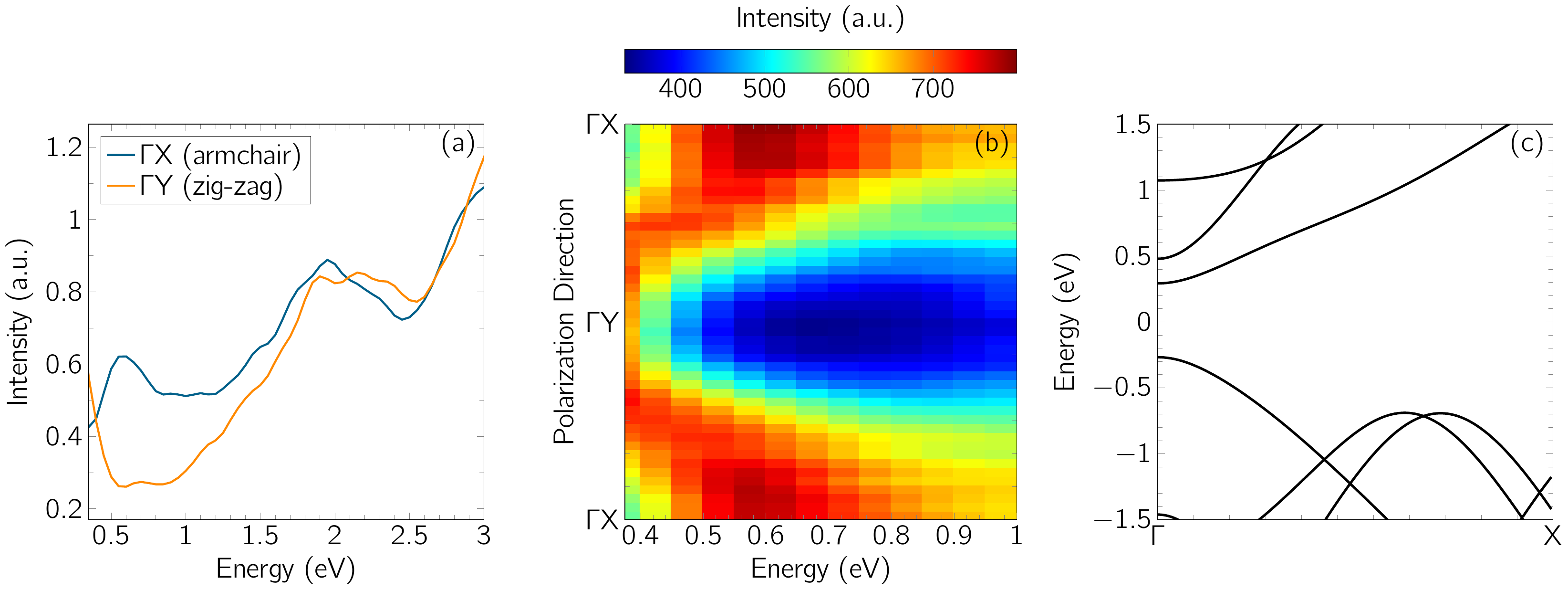}
  \caption{(Color online) (a) EELS intensity along the two high-symmetry directions within the puckered layers of \bp. (b) Low-energy polarization-map showing the smooth angular dependence of the EELS intensity within the puckered layers of \bp. (c) Calculated single-particle band structure along \GX. Data in (a) and (b) have been measured with a fixed momentum transfer of $q=\rangstrom{0.1}$.}
  \label{fig:pol_dependence}
\end{figure*}

From \figref{fig:pol_dependence} it is clear that this translates directly to a strongly anisotropic EELS response. Along the \GX (armchair) direction we find a low-energy mode that resides at about $E=0.6\unit{eV}$, slightly above the gap edge of $E_G=0.3\unit{eV}$ \cite{Akahama_phys.stat.sol.b_2001_v223_p349}. When changing the polarization direction from the \GX- to the \GY (zig-zag)-direction within the puckered layers, this mode gradually shifts to lower energy and loses strength as can be seen from  \figref{fig:pol_dependence}(b).  

Considering a simple Lorentz picture for the observed structure it is important to note that a peak in the EELS response occurs not at the energy $\omega_0$ of the electronic transition but at $\tilde{\omega}=\sqrt{\omega_0^2+\omega_p^2}$ where $\omega_p$ is a measure of the spectral weight acquired by this excitation \cite{Kuzmany_SolidStateSpectroscopy_2009}. It can be shown that in the limit of $\epsilon_\infty\rightarrow1$, where $\epsilon_\infty$ is the background dielectric constant, $L(\tilde{\omega})\sim\omega_p/\omega_0$ which means that a concomitant lowering of intensity and energetic position of a transition observed in EELS can only be reconciled by a diminished spectral weight $\omega_p$. Consequently, we attribute the behavior of the measured EELS intensity shown in \figref{fig:pol_dependence} to an angular dependent reduction of spectral weight when changing the polarization from \GX to \GY.

\begin{figure*}[ht]
  \centering
\includegraphics[scale=.5]{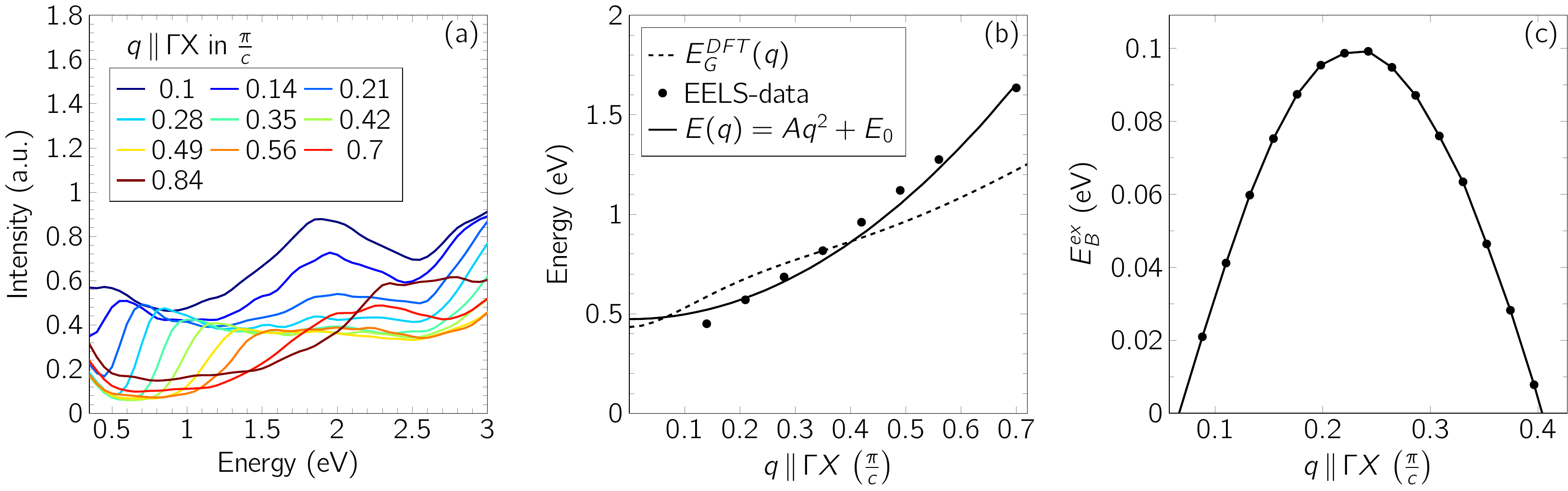}
  \caption{(Color online) (a) The momentum evolution of the measured EELS intensity for momentum transfers along the armchair direction. (b) the dispersion of the low-energy feature as extracted from the low-energy edges of the measured intensities. The solid line corresponds to a quadratic fit with the values: $A=2.412~\text{eV}\,\text{\AA}^{2}$ and $E_0=0.473\unit{eV}$ and the dashed line to the single-particle gap derived from the band-structure calculation shown in \figref{fig:pol_dependence}. (c) The evolution of the exciton binding energy $E_B^{ex}$ as a function of momentum transfer along the armchair direction.}
  \label{fig:armchair_dispersion}
\end{figure*}

Moving away from the center of the BZ this low-energy mode shows a strong momentum dependence (see \figref{fig:armchair_dispersion}(a)) which hints at a highly dispersive and therefore delocalized excitation. This is further corroborated by \figref{fig:armchair_dispersion}(b) where we show the dispersion of this feature as evaluated by tracking the steepest slope on the low-energy side. We emphasize that along the \GY direction there is no dispersion and the spectrum remains largely unaffected in the corresponding energy window when increasing the momentum transfer. In addition, \figref{fig:armchair_dispersion}(b) contains the size of the single-particle gap $E_G^{DFT}(q)=E_{CB}(q)-E_{VB}(\Gamma)$ as derived from our DFT calculations.

As can be seen, the feature under investigations shows a bandwidth of more than $1.5\unit{eV}$ within the accessible momentum window. %Note that the edge of the BZ lies at about $\rangstrom{0.72}$.
More importantly, when fitted to a quadratic function, its onset  $E_0=E(q=0)=0.473\unit{eV}$ lies in close proximity to the single-particle gap.  Note that we cannot measure at lower $q$-values because of the increasing contribution of surface scattering for $q\rightarrow0$ \cite{Fink_Adv.Electron.ElectronPhys._1989}. Comparing the bandgap extracted from our DFT calculations to the measured EELS data, there is clear evidence that the dispersion of the low-energy mode seen in EELS lies \emph{below} the single-particle gap at finite $q$ which is suggestive of a bound, i.e. excitonic state. 
The fit to our data shown in \figref{fig:armchair_dispersion}(b) yields an effective exciton mass of $M^\ast=\hbar^2/2A\sim 1.6\,m_e$ with the free-electron mass $m_e$. This is in reasonable agreement with the value of $M^\ast_{DFT}(\Gamma)=m^\ast_h(\Gamma)+m^\ast_e(\Gamma)\sim1.1\,m_e$ obtained within the effective-mass approximation for the exciton from our DFT calculations. As shown in \figref{fig:armchair_dispersion}(c), the binding energy defined as $E_B^{ex}=E_G^{DFT}(q)-E(q)$ of this exciton %monotonically increases as a function of momentum. 
has a well pronounced maximum of about $0.1\unit{eV}$ for $q_{max}\approx 0.23~\pi/c$.
We emphasize that as soon as $E_B^{ex}\leq0$ %in the limit $q\rightarrow0$ 
the exciton ceases to exist as it will become over-damped by single-particle excitations. As the exciton binding energy $E_B^{ex}(q)\sim 1/\epsilon(q)$ and $\epsilon(q)$ strongly decreases for finite $q$ \cite{Penn_Phys.Rev._1962_v128_p2093,Levine_Phys.Rev.B_1982_v25_p6310,Resta_Phys.Rev.B_1977_v16_p2717} the increase of $E_B^{ex}(q)$ for $q<q_{max}$ seen in \figref{fig:armchair_dispersion}(c) is a result of the less effective screening of the Coulomb interaction for smaller wavelengths. This effect is, however, overcompensated by the merging of the exciton dispersion with the single particle continuum which results in the decay of the exciton into unbound electron-hole pairs, thereby producing an increasing broadening of the spectral features for higher momentum values.

In order to obtain the full optical response, the measured EELS data have been corrected for contributions from the quasi-elastic line and multiple-scattering. Subsequently, the signal was subject to a KKA and normalized to yield $\epsilon_1(\omega=0)$ according to \cite{Nagahama_J.Phys.Soc.Jpn._1985_v54_p2096}.

\begin{figure*}%[ht]
  \begin{subfigure}[]{.4\textwidth}
    \includegraphics[scale=.75]{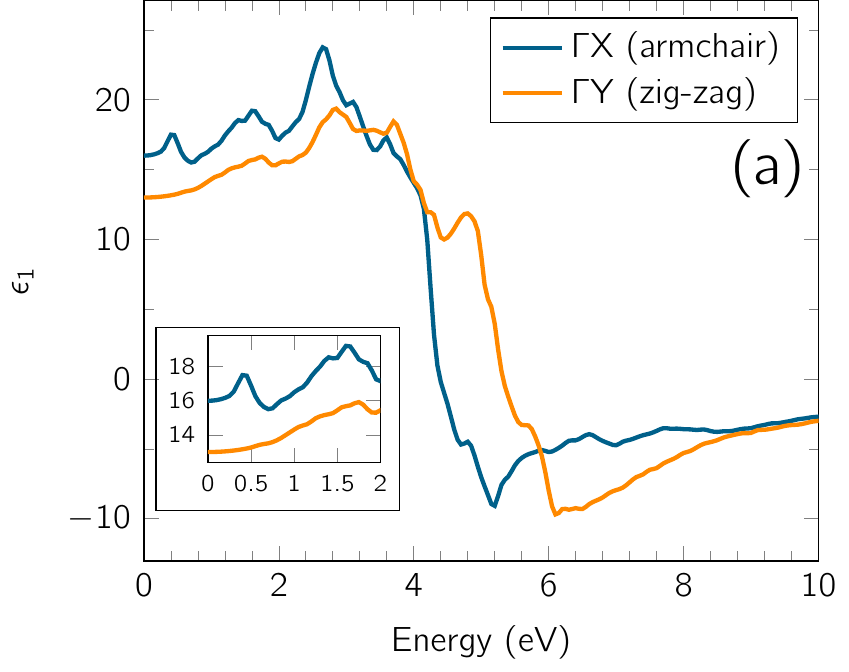}
  \end{subfigure}
  \begin{subfigure}[]{.4\textwidth}
    \includegraphics[scale=.75]{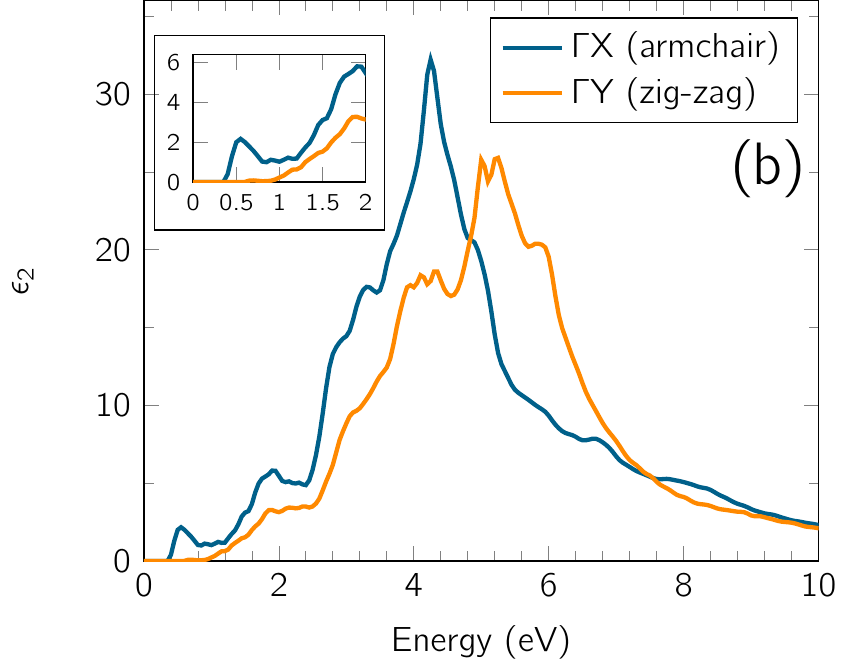}
  \end{subfigure}
  \caption{(Color online) $\epsilon_1$ (a) and $\epsilon_2$ (b) along the two high-symmetry directions within the puckered layers of \bp as derived from a KKA of the corresponding EELS data. The insets show the low-energy region.}
  \label{fig:e1_e2}
\end{figure*}

The resulting curves for the real and imaginary parts of $\epsilon(\omega)$ are shown in \figref{fig:e1_e2}. As expected already from the plain EELS data shown in \figref{fig:pol_dependence}(a), along the armchair direction there is a well pronounced mode right above the bandgap which is invisible for polarizations along the zig-zag direction. %The verification of the experimental $E_G$ in $\epsilon_2(\omega)$ shows the reliability of our KKA. In addition, 
The high values of $\epsilon_1$ seen for small momentum transfers in \figref{fig:e1_e2}(a), which inevitably have to decrease for $q>0$ \cite{Penn_Phys.Rev._1962_v128_p2093,Levine_Phys.Rev.B_1982_v25_p6310,Resta_Phys.Rev.B_1977_v16_p2717}, support our interpretation of an increasing $E_B^{ex}$ (see \figref{fig:armchair_dispersion}(c)) in terms of weaker screening at least for $q<q_{max}$.
 
Strongly anisotropic optical properties have also been predicted for phosphorene by \emph{ab-initio} calculations \cite{Tran_Phys.Rev.B_2014_v89_p235319,Qiao_NatCommun_2014_v5_p4475}. In particular, it has been shown that phosphorene hosts an excitonic state  which is strongly delocalized along the armchair direction but confined to a small region along the zig-zag direction \cite{Tran_Phys.Rev.B_2014_v89_p235319,Rodin_Phys.Rev.B_2014_v90_p75429}. This is consistent with our interpretation given above for the polarization dependence of the EELS signal in the bulk. The large extension of the exciton along the armchair direction produces a large dipole moment and hence spectral weight along this direction, yielding the peak seen in \figref{fig:pol_dependence}(a) and (b). Conversely, changing the polarization towards \GY will progressively project out a smaller dipole moment  that lowers the intensity (and in EELS also energy) due to the small extension of the exciton along the zig-zag direction.

Lowering the dimensionality from the bulk to the two-dimensional phosphorene will cause smaller interlayer screening resulting in a larger value of the single-particle gap \cite{Tran_Phys.Rev.B_2014_v89_p235319,Qiao_NatCommun_2014_v5_p4475} as well as an enhanced excitonic binding energy due to the dependence of the dielectric function on the gap size \cite{Penn_Phys.Rev._1962_v128_p2093,Levine_Phys.Rev.B_1982_v25_p6310}. Recent theoretical approaches predict excitonic binding energies of about $0.8\unit{eV}$ in monolayer phosphorene \cite{Rodin_Phys.Rev.B_2014_v90_p75429,Tran_Phys.Rev.B_2014_v89_p235319}. Therefore, in phosphorene, the exciton might eventually be pushed out of the single-particle continuum already at the $\Gamma$-point and produce strong optical absorption along the armchair direction. This emergence of the exciton out of the continuum should then be observable as a function of layer-thickness by conventional optical means. Indeed, recent photoluminescence data strongly argue in favor of layer-dependent excitonic absorption \cite{Zhang_ACSNano_2014_v8_p9590}.

From investigations on organic semiconductors it is well-known that delocalized, i.e. strongly dispersing  excitonic modes contribute significantly to the mobility-values via the exciton diffusion coefficient \cite{Cheng_J.Chem.Phys._2003_v118_p3764,Yamagata_J.Chem.Phys._2011_v134_p204703}. Our data therefore may provide a route to reconcile the high mobility values and their anisotropy predicted and reported in transport measurements on phosphorene \cite{Li_NatNano_2014_v9_p372,Qiao_NatCommun_2014_v5_p4475,Xia_NatCommun_2014_v5_p4458,Liu_ACSNano_2014_v8_p4033,Cai_ScientificReports_2014_v4_p6677,Zhang_ACSNano_2014_v8_p9590} with the microscopic fate of the excitonic states in this monolayer material.

%\section{Summary}
\label{sec:summary}
In summary, we investigated the energy- and momentum-dependent optical response in the bulk of black phosphorus. We find strong anisotropies within the puckered layers with the armchair direction showing much stronger  absorption in the infrared and visible part of the spectrum. In addition we identified an excitonic mode for finite values of momentum transfer with an effective mass of about $1.6\,m_e$. This mode can be probed in phosphorene by conventional optical spectroscopy and might be interesting for future optoelectronic applications. 

%\section{Acknowledgments}
We highly appreciate technical support from R. Hübel, S. Leger and M. Naumann.
\bibliographystyle{apsrev4-1}
\bibliography{bp_manuscript}
%\listoftodos
\end{document}